\documentclass[%
	prl,floatfix, superscriptaddress,amsmath,amssymb,
	reprint,
]{revtex4-1}
\usepackage{float}
\usepackage[english]{babel}
\usepackage [autostyle, english = american]{csquotes}
\MakeOuterQuote{"}
\usepackage{amsmath}
\usepackage{amsthm}
\usepackage{mathtools}
\usepackage{dsfont}
\usepackage{amssymb}
\usepackage{hyperref}
\usepackage{color}
\usepackage{xcolor}
\usepackage{subcaption}
\usepackage{standalone}
\usepackage{tikz}
\usetikzlibrary{decorations.pathreplacing,calligraphy}
\usetikzlibrary{arrows.meta,arrows}
\usepackage{silence}
\usepackage{braket}
\usepackage{graphicx}
\usepackage{dcolumn}
\usepackage{bm}
\renewcommand{\selectlanguage}[1]{}
\definecolor{natgreen}{HTML}{7BB725}
\definecolor{natgrey}{HTML}{c7c7c7}
\usepackage[font=small,labelfont=bf,
   justification=Justified,
   format=plain]{caption}

\begin{document}
\preprint{AIP/123-QED}

\title{Quantum Convolutional Neural Network for Phase Recognition in Two Dimensions}

\author{Leon C. Sander}
\affiliation{Department of Physics, Friedrich-Alexander University Erlangen-Nürnberg
	(FAU), Erlangen, Germany}
 
\author{Nathan A. McMahon}
\affiliation{Department of Physics, Friedrich-Alexander University Erlangen-Nürnberg
	(FAU), Erlangen, Germany}
\affiliation{$\langle aQa^{L}\rangle$ Applied Quantum Algorithms, Leiden University, Netherlands}
\affiliation{LIACS, Leiden University, Netherlands}
 
\author{Petr Zapletal}
\affiliation{Department of Physics, Friedrich-Alexander University Erlangen-Nürnberg
	(FAU), Erlangen, Germany}
\affiliation{Department of Physics, University of Basel, Basel, Switzerland}

\author{Michael J. Hartmann}
\affiliation{Department of Physics, Friedrich-Alexander University Erlangen-Nürnberg
	(FAU), Erlangen, Germany}

\date{\today}

\begin{abstract}
     Quantum convolutional neural networks (QCNNs) are quantum circuits  for characterizing complex quantum states. They have been proposed for recognizing quantum phases of matter at low sampling cost and have been designed for condensed matter systems in one dimension. Here we construct a QCNN that can perform phase recognition in two dimensions and correctly identify the phase transition from a Toric Code phase with $\mathbb{Z}_2$-topological order to the paramagnetic phase. The network also exhibits a noise threshold up to which the topological order is recognized. Furthermore, it captures correlations between all stabilizer elements of the Toric Code, which cannot be accessed by direct measurements. This increases the threshold for errors leading to such correlations and allows for correctly identifying the topological phase in the presence of strong correlated errors. Our work generalizes phase recognition with QCNNs to higher spatial dimensions and intrinsic topological order, where exploration and characterization via classical numerics become challenging. 
\end{abstract}

\maketitle

With the remarkable progress in quantum computing technologies \cite{aruteQuantumSupremacyUsing2019a,krinnerRealizingRepeatedQuantum2022,postlerDemonstrationFaulttolerantUniversal2022,bluvsteinLogicalQuantumProcessor2023b}, the output states of quantum computers and simulations are becoming too complex to characterize them with direct measurements and classical computing. Although many local properties of quantum states can be efficiently characterized using shadow tomography \cite{huangPredictingManyProperties2020}, the characterization of global properties remains an open challenge. Direct processing of quantum data on quantum computers offers a potential solution to this problem, where recent developments include quantum autoencoders \cite{romeroQuantumAutoencodersEfficient2017,bondarenkoQuantumAutoencodersDenoise2020}, quantum principle component analysis \cite{lloydQuantumPrincipalComponent2014}, certification of Hamiltonian dynamics \cite{wiebeHamiltonianLearningCertification2014, gentileLearningModelsQuantum2021}, and quantum reservoir processing \cite{ghoshQuantumReservoirProcessing2019}. Particularly promising are quantum neural networks that use parameterized quantum circuits, measurement and feed-forward \footnote{Feed-forward refers to a quantum operation conditioned on the outcome of a projective mid-circuit measurement, typically used in quantum error correction.} to characterize large amounts of quantum data \cite{farhiClassificationQuantumNeural2018, congQuantumConvolutionalNeural2019, beerTrainingDeepQuantum2020, kottmannVariationalQuantumAnomaly2021, gongQuantumNeuronalSensing2023}.

A central field for applications of quantum computing is condensed matter physics, where the characterization of quantum states is of central importance. In particular, the classification of phases of matter \cite{kitaevFaulttolerantQuantumComputation2003,savaryQuantumSpinLiquids2016} is a key requirement for understanding strongly correlated materials \cite{sachdevQuantumPhaseTransitions1999}.
While phases of matter in one-dimensional systems are well understood thanks to accurate classical numerical simulations \cite{mccullochInfiniteSizeDensity2008}, two- and higher-dimensional systems are notoriously hard to simulate on classical computers \cite{zhengStripeOrderUnderdoped2017, hangleiterEasingMonteCarlo2020}. 
Of particular interest in this context are topological many-body phases of matter, which have already been prepared on quantum computers  \cite{satzingerRealizingTopologicallyOrdered2021,semeghiniProbingTopologicalSpin2021,bluvsteinQuantumProcessorBased2022,iqbalTopologicalOrderMeasurements2023}, thus opening new avenues for their exploration. However, the characterization of these topological phases remains challenging because of the absence of local order parameters. Other characteristic quantities, such as fidelity susceptibility and entanglement, are hard to measure. Previous approaches addressing this challenge include applying classical machine learning techniques to data from classical numerical simulations \cite{vannieuwenburgLearningPhaseTransitions2017, carrasquillaMachineLearningPhases2017,greplovaUnsupervisedIdentificationTopological2020} and classical shadows \cite{huangProvablyEfficientMachine2021}. These approaches are however limited by the large amounts of data required for describing quantum states on classical computers. 

Quantum convolutional neural networks (QCNNs) have been proposed to substantially reduce the sample complexity of recognizing symmetry-protected topological phases in one dimension \cite{congQuantumConvolutionalNeural2019,lakeExactQuantumAlgorithms2022} and were shown to uncover characteristic properties of such phases from training data via supervised learning \cite{congQuantumConvolutionalNeural2019, caroGeneralizationQuantumMachine2022,liuModelIndependentLearningQuantum2022}. Moreover, these QCNNs can be explicitly constructed by combining techniques from quantum error correction and renormalization-group flow \cite{congQuantumConvolutionalNeural2019,lakeExactQuantumAlgorithms2022,herrmannRealizingQuantumConvolutional2022,zapletalErrortolerantQuantumConvolutional2024}, which is a well-established method for the classification of phases of matter \cite{sachdevQuantumPhaseTransitions1999}. Recently, QCNN s which are robust against errors in quantum data \cite{zapletalErrortolerantQuantumConvolutional2024}, were implemented on a superconducting quantum processor under NISQ conditions \cite{herrmannRealizingQuantumConvolutional2022}. 

So far, QCNNs have, however, only been applied to one-dimensional systems, which can be accurately simulated on classical numerical computers. Their extension to two- and higher-dimensional systems, where quantum computers can potentially provide a computational advantage, has remained largely unexplored.  Existing work includes a method for detecting topological phases in two-dimensions called Locally Error-corrected Decoration (LED) \cite{congEnhancingDetectionTopological2024}, which is inspired by QCNNs. 
However, all current results are limited to snapshot-based LED, i.e. the classical post-processing of direct measurements. Thus, in its current form, it faces limitations once the characterization of the quantum states requires more data than can be efficiently processed on classical computers. This can occur at interesting points, as for example close to the critical point of a quantum phase transition. Hence extending QCNNs to two dimensions is a non-trivial generalization as quantum phases in two- and higher-dimensional systems are fundamentally different due to long-range entanglement, giving rise to topological order, which is absent in one dimension \cite{wenColloquiumZooQuantumtopological2017}.

Here, we address this open challenge by generalizing QCNNs to detect $\mathbb{Z}_2$-topological order in two-dimensional spin arrays. We analytically construct a QCNN inspired by quantum error correction and  multiscale entanglement renormalization ansatz \cite{Vidal_Mera_2008}. Using matrix-product-state simulations, we show that the QCNN recognizes the topological phase of
the Toric Code Hamiltonian in a magnetic field from a topologically trivial phase. In contrast to existing methods \cite{congEnhancingDetectionTopological2024}, which struggle to correctly classify complex quantum states close to phase boundaries, the QCNN precisely identifies the critical value of the magnetic field where the Toric Code undergoes the phase transition. Moreover, we show that the QCNN output is robust against incoherent errors below a threshold error probability allowing for the application of the QCNN on current and near-term quantum computers.

Classical approaches based on direct measurements, such as the snapshot-based LED method in \cite{congEnhancingDetectionTopological2024}, cannot fully capture quantum correlations. In the case of the Toric Code, the stabilizer elements would require readout in different Pauli bases and thus cannot be simultaneously measured via direct measurements on typical quantum processors. 
In contrast, in processing quantum states via the QCNN quantum circuit, correlations between all stabilizer elements can be taken into account, increasing the threshold probability for errors that lead to such correlations, similarly to correlated quantum error correction \cite{fowlerOptimalComplexityCorrection2013}.
As a result, the QCNN correctly identifies the topological phase in the presence of strong correlated errors, where classical approaches misclassify topological states. Note that, in contrast to quantum error correction constructions \cite{krinnerRealizingRepeatedQuantum2022}, it does not require additional readout qubits. 

The QCNN we consider is a quantum circuit that can be written as a unitary 
\begin{equation}
	U_\mathrm{QCNN}=U_\mathrm{FC} U_\mathrm{P}^{(d)}...U_\mathrm{P}^{(2)}
	U_\mathrm{P}^{(1)} U_\mathrm{C}\,,
 \label{eqQCNN}
\end{equation}
consisting of a convolutional layer $U_\mathrm{C}$, $d$ pooling layers $U_\mathrm{P}^{(l)}$, for $l=1,2,...,d$, and a fully connected layer $U_\mathrm{FC}$ \cite{
herrmannRealizingQuantumConvolutional2022,zapletalErrortolerantQuantumConvolutional2023}. The processing of quantum states in the QCNN is schematically depicted in \autoref{QCNN_structure}. The convolutional layer $U_\mathrm{C}$ performs a translationally invariant transformation. In each pooling layer, the system size is reduced by discarding a fraction of qubits after applying gates on the remaining qubits. The fully connected layer $U_\mathrm{FC}$ prepares the output state of the remaining qubits, which are subsequently measured.

We construct the convolutional layer from Clifford gates and the pooling layers from single-qubit Pauli gates controlled by qubits that are subsequently discarded. In this case, the QCNN circuit \autoref{eqQCNN} featuring only the single convolutional layer is equivalent to a deeper structure proposed in Ref.~\cite{congQuantumConvolutionalNeural2019}, consisting of $d$ interleaved convolutional and pooling layers \cite{herrmannRealizingQuantumConvolutional2022,zapletalErrortolerantQuantumConvolutional2023}. The QCNN thus performs the same unitary with a substantially shorter circuit, see the supplemental material \cite{SM} for details. Moreover, all such pooling layers can be efficiently performed in classical post-processing and only the initial convolutional layer $U_\mathrm{C}$ is executed on the quantum device.

\begin{figure}
        \includegraphics[width=1\columnwidth]{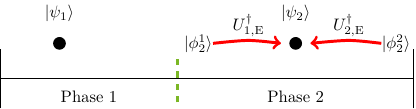}
	\caption{Schematic of phase recognition via a QCNN for two phases with characteristic states $\ket{\psi_1}$ and $\ket{\psi_2}$. Other states $\ket{\phi_j^k}$ in the respective phases can be generated by local unitary perturbations $\ket{\phi_j^k}=U_{k,\mathrm{E}}\ket{\psi_j}$ and the phase can, for each input state, be identified by undoing the respective perturbation $U_{k,\mathrm{E}}$ via sufficient error correction. } 
	\label{RGFlow}
\end{figure}

\begin{figure}[t]
 \includegraphics[width=\columnwidth]{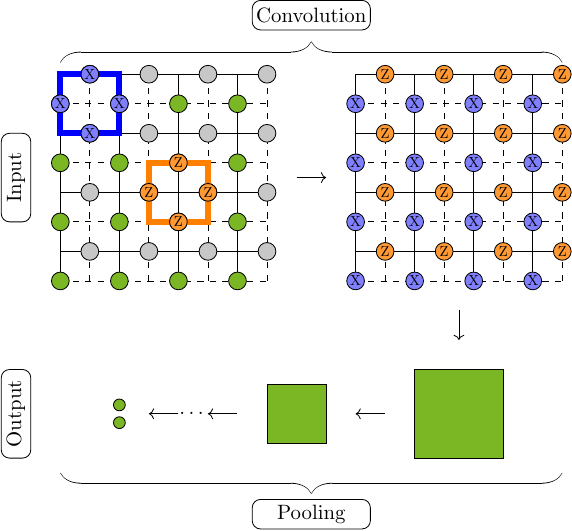}
	\caption{QCNN structure. The first lattice on the input side shows the Toric Code lattice with dimension $d=4\times 4$, which is periodic in vertical and horizontal direction. The qubits on the respective sublattices are represented by gray (horizontal edges) and green circles (vertical edges). Examples of plaquette ($A_p$) and vertex ($B_v$) operators are highlighted in blue and orange. The circuit of the convolution maps the stabilizers of the input lattice to measurements on individual qubits of the respective colors, i.e.,  blue (green) qubits carry the measurement of the plaquette (vertex) stabilizer that is anchored on the qubits to their left. This convolution evolves error patterns on the input state and brings them into a form that can be corrected by the local operations in the pooling if the input was in the topological phase. Iterated pooling procedures reduce the lattice dimension (green squares) with every step until only two output qubits are left (green dots) and read out.}
	\label{QCNN_structure}
\end{figure}

The functionality of our QCNN is based on the fact that topological order is robust against local perturbations \cite{kitaevFaulttolerantQuantumComputation2003}. Hence, one can define characteristic states $\ket{\psi_i}$ for each phase of the system and treat other states of the same phase as characteristic states masked by unitary
perturbations $\ket{\phi_j^k}=U_{k,\mathrm{E}}\ket{\psi_j}$  \cite{lavasaniStabilityGappedQuantum2024}, which we identify as errors. If one can recover the characteristic state by removing
these errors, one has successfully identified the corresponding phase. In
\autoref{RGFlow} we sketch such a phase space and the recovery of the reference
state. The explicit structure of our QCNN can be found in the supplemental material \cite{SM}.

At each layer $l$, we can generate an output by `measuring' the qubits in the computational basis,
\begin{equation}
	M^l_\mathrm{QCNN}=\dfrac{1}{N_l}\sum_{j=1}^{N_l} m_j, \,
	\label{QCNN_output}
\end{equation}
where $m_j\in \{-1,1\}$ is
the measurement outcome of qubit $j$ and $N_l$ the number of qubits in layer
$l$. The measurement outcomes $m_j$ correspond to the values of plaquette and star operators on a lattice with the reduced system size $N_l$. The output $M^l_\mathrm{QCNN}$ thus measures how close the output state is to the error-free ground state of the Toric Code after having performed $l$ pooling layers.

As the pooling layers are performed in classical post-processing, their controlled Pauli gates are applied classically on the bit string of computational basis measurements for all qubits and the 'measurements' correspond to simply recording the classical data.
The support of the operator $M^l_\mathrm{QCNN}$ as defined in \autoref{QCNN_output}, will grow exponentially with the number of layers $l$. For our QCNN, we will obtain the following averages over a large number of samples
\begin{equation}
	\braket{\mathrm{QCNN}} \coloneqq \braket{M^d_\mathrm{QCNN}}=
	\begin{cases}
		1   & \text{if topological phase}  \\
		0   & \text{if non-topological phase}\,.
	\end{cases}
\end{equation}

We focus on the two-dimensional Toric Code \cite{Kitaev06} with periodic boundary
conditions. It is a natural candidate for exploring QCNNs for two-dimensional systems as it shows intrinsic $\mathbb{Z}_2$-topological order \cite{Kitaev06}. The Toric Code is a two-dimensional square lattice with the qubits located on its edges. For exploring its condensed matter physics, it can be described by the Hamiltonian
\begin{equation}
	\label{eqn:Toric_pure}
	H_{\mathrm{TC}}=-\sum_p A_p -\sum_v B_v ,
\end{equation}
with the plaquette operators $A_p =\prod_{i=1} ^4 X_i$ and the star
operators $B_v =\prod_{i=1} ^4 Z_i$, where $X_{i}$ ($Z_{i}$) is the Pauli-X(Z) operator on qubit $i$, see
\autoref{QCNN_structure}. 
Exposing the Toric Code to an external magnetic field gives rise to a rich phase diagram featuring the topologically ordered phase and a trivial phase, separated by first- or second-order phase transitions depending on the orientation of the field \cite{trebstBreakdownTopologicalPhase2007, Dusuel11}. To explore these regimes, we modify \autoref{eqn:Toric_pure} to include applied magnetic fields in X/Y/Z-direction with strengths $h_X$, $h_Z$ and $h_Y$,
\begin{equation}
	H_{\mathrm{mag}}= H_{\mathrm{TC}} - h_Z \sum_i ^N Z_i -h_X \sum_i ^N X_i-h_Y \sum_i ^N Y_i\,.
	\label{eq:toric_mag}
\end{equation}
The QCNN that we apply to (noisy) ground states of this model is sketched in \autoref{QCNN_structure} and its detailed description can be found in the supplemental material \cite{SM}.
The QCNN we here propose is capable of detecting the topologically ordered phase with high accuracy. This means that the QCNN's output converges to 1 with increasing depth $d$ if the input state is in the topological phase 
and to 0 otherwise.

To evaluate its performance, we carried out several numerical simulations, where we use Matrix Product State (MPS) simulations \cite{vidalClassicalSimulationInfiniteSize2007} to approximate ground states for large lattices \autoref{eq:toric_mag} and classical simulations of Clifford gates on Pauli strings to explore the effects of incoherent Pauli noise. We show the technical details of our computations in the supplemental material \cite{SM}. 

\begin{figure}
	\centering
	\includegraphics[width=\columnwidth]{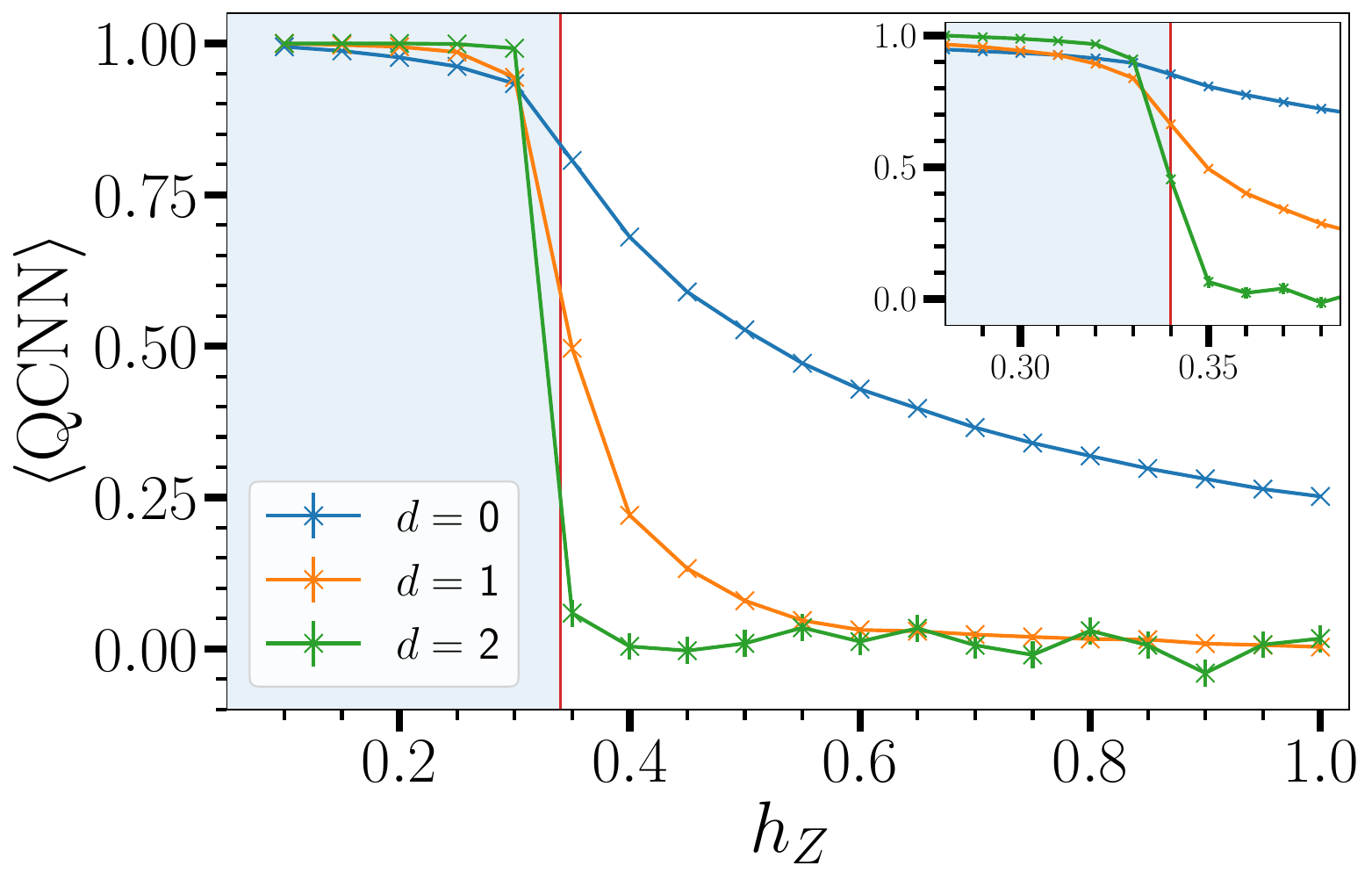}
	\caption{QCNN output for a variation of the magnetic field strength $h_Z$, where $h_X = 0$, and for different depths $d$ of the QCNN, corresponding to the number of pooling layers $U_\mathrm{P}^{(l)}$. Each data point represents 2000 samples from the MPS representation of the corresponding ground state with the bond dimension $\chi=1250$ for an infinite cylinder with the periodic dimension $l_2 = 9$. The shaded region corresponds to the topologically ordered phase and the vertical line to critical magnetic field strength $h^*_Z=0.34$. The inset shows the output for different samples with higher resolution around the phase transition, which are calculated with a bond dimension of $\chi=1500$. For all samples, we observe increased steepness at the phase transition in the QCNN output with increasing QCNN depth corresponding to greater precision in the phase recognition.}
	\label{DMRG_phase_diagram}
\end{figure}

We first test the QCNN on ground states of $H_\textrm{mag}$ with a magnetic field in Z-direction, $h_X = h_Y = 0$ in \autoref{eq:toric_mag}.  \autoref{DMRG_phase_diagram} shows the resulting phase diagram, where we consider field strengths $0 \leq h_Z \leq 1$. From iDMRG simulations for our cylindrical lattice, we determine the phase boundary at $h_Z^*= 0.34$ as a peak in the second derivative $d^{2}E/dh_{Z}^{2}$ of the ground state energy with respect to $h_Z$, which we mark in the phase diagram. The plot of the QCNN output $\braket{M^l_\mathrm{QCNN}}$, see \autoref{QCNN_output}, shows a transition at this expected point up to the sample resolution. Additionally, the sharpness of the transition increases with the layer number in the sense that the absolute value of the gradient of $\braket{M^l_\mathrm{QCNN}}$ increases, which makes the transition in the output more step-like with an increasing number of pooling layers, thus showing the successful recognition of the topological phase by the QCNN.

Furthermore, we test whether our QCNN is able to recognize the topological phase in the presence of incoherent noise. We do this by considering the Toric Code ground state perturbed by Pauli noise. Specifically, we apply random Pauli noise to each individual qubit in the form of the error channel
\begin{equation}
\rho=\mathcal{E}(\ket{\psi}\bra{\psi})=\sum_{l=0}^m K_l \ket{\psi}\bra{\psi}K_l^\dagger\,,
    \label{PauliNoise}
\end{equation}
with $K_l\in\{\sqrt{p_\mathds{1}}\mathds{1},\,\sqrt{p_X}X,\,\sqrt{p_Y}Y,\sqrt{p_{Z}}Z\}^{\otimes N}$ to the input qubits, where $p_{E}$ are probabilities of Pauli errors $E=X,Y,Z$ and $p_\mathds{1}=1- p_{X} - p_Y -p_{Z}$. As the QCNN only consists of Clifford and controlled Pauli gates, we can then subsequently track the evolution of the Pauli error strings on the lattice under the operations of the QCNN. It is sufficient to employ the following gate identities $X_j=H_j Z_j H_j$, $\textrm{CNOT}_{i,j}X_i =X_i X_j \textrm{CNOT}_{i,j}$ and $\textrm{CNOT}_{i,j}Z_j =Z_i Z_j \textrm{CNOT}_{i,j}$,
where $H_j$ is the Hadamard gate on qubit $j$ and $\textrm{CNOT}_{i,j}$ applies a Pauli $X$ on qubit $j$
conditioned on qubit $i$. 

\begin{figure}
	\centering
	\includegraphics[width=\columnwidth]{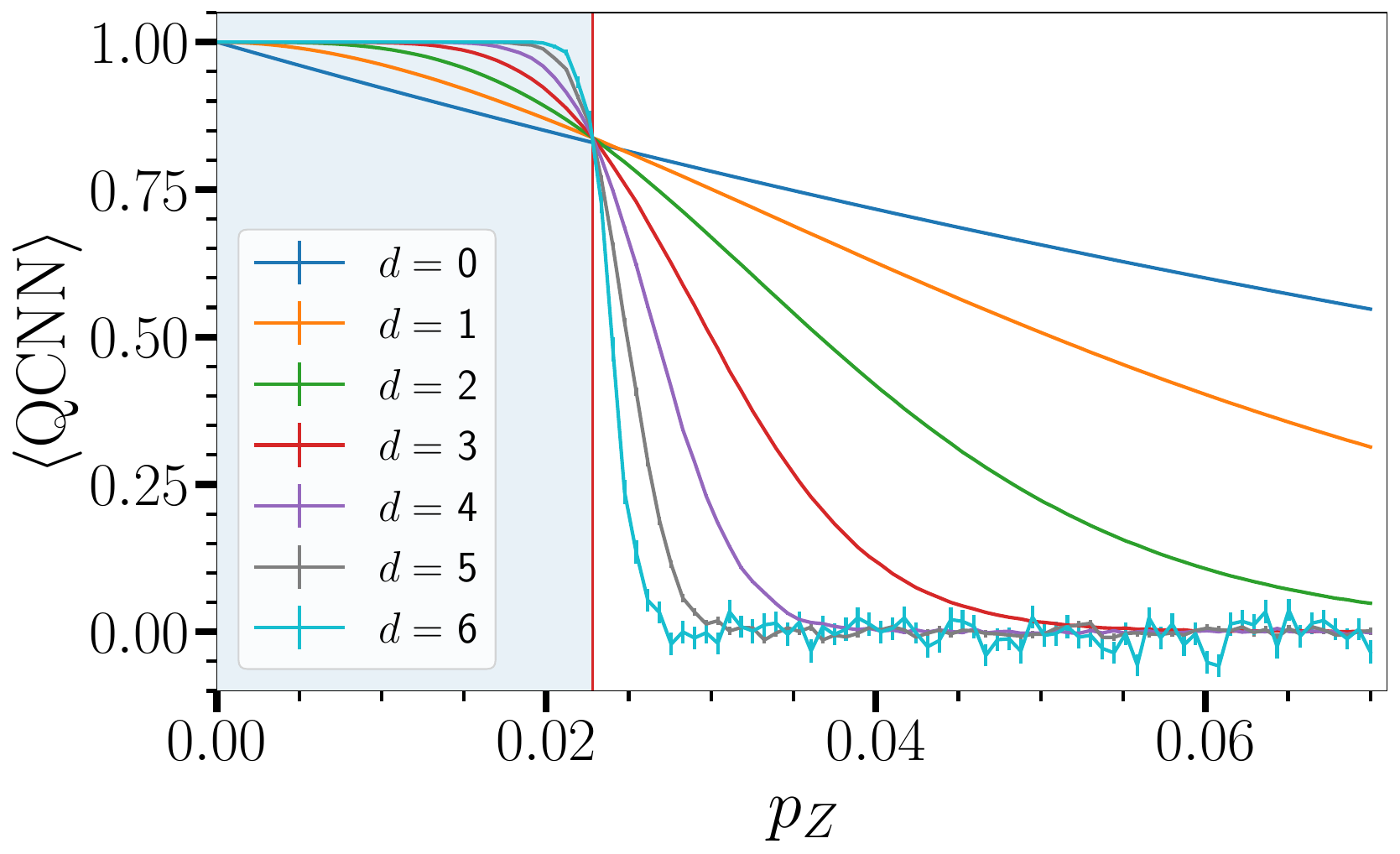}
	\caption{QCNN output for incoherent Pauli noise perturbing the Toric Code ground state as in \autoref{PauliNoise} with $p_X=p_Y=0$. Via tracking the evolution of Pauli strings on the lattice under the convolution and the pooling layers we can classically simulate the QCNN output for large system sizes. For this plot, we calculate the results for about 9.6 Million qubits, which allows us to employ more pooling layers compared to the MPS samples. As for the case of the magnetic field, we find a transition in the QCNN output that becomes more pronounced with increasing depth of the QCNN. The red vertical line marks the error threshold of $p_\mathrm{th}=2.28\%$.}
	\label{7laysOutput}
\end{figure}

We present results for the QCNN output, \autoref{QCNN_output}, for cases where the input state is the reference state of the topological phase, i.e. a ground state of $H_{\mathrm{TC}}$ as in \autoref{eqn:Toric_pure}, perturbed by Pauli-Z noise ($p_X=p_Y=0$) in \autoref{7laysOutput}, which shows a transition in the QCNN output over the Pauli-Z error probability $p_Z$. The QCNN output converges to the ideal noise-free value $\braket{M^d_\mathrm{QCNN}}=1$ with the increasing number of pooling layers $d$ below the threshold error probability $p_{\rm th} = 2.28\%$ (red vertical line). On the other hand, the output decreases to zero above this threshold. In the supplemental material \cite{SM}, we also show results for simultaneous Pauli-Z and Pauli-X noise, and demonstrate that the QCNN detects the phase transition at the critical strength $h_Z^*$ of the magnetic field in the presence of noise.

We now discuss the key difference between the QCNN, and approaches based on direct measurements and snapshot-based processing as in \cite{congEnhancingDetectionTopological2024}. On most quantum processors, only measurements in local Pauli bases can be directly performed. In this case, plaquette operators $A_p$ and star operators $B_v$ cannot be simultaneously measured as they require readout in different Pauli bases. As a result, such direct measurements do not capture the correlations between these operators and only uncorrelated pooling, where the measurement outcomes of plaquette and star operators are processed separately, can be performed. In our QCNN, in contrast, all $A_p$ and $B_v$ operators are simultaneously measured after having applied the convolutional layer as a quantum circuit, see \autoref{QCNN_structure}. Therefore, we can implement correlated pooling that takes into account correlations between these operators similarly to correlated quantum error correction \cite{fowlerOptimalComplexityCorrection2013}. See the supplemental material \cite{SM} for details about correlated pooling and its numerical simulation.

\begin{figure*}
    \centering
    \includegraphics[width=2\columnwidth]{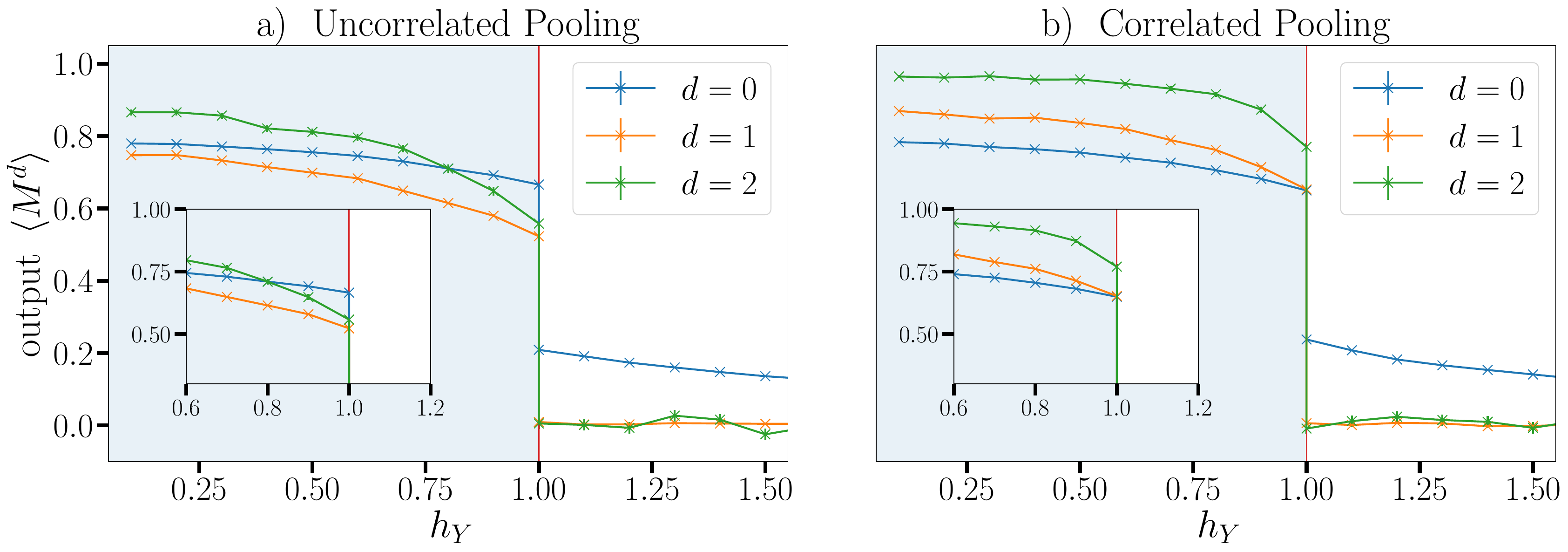}
    \caption{
    Output of (a) uncorrelated pooling and (b) correlated pooling for the magnetic field strength $h_Y$ in the presence of Pauli-Y noise with $p_Y=3\%$ and for different depths $d$. Each data point represents 2000 samples from the MPS representation of the corresponding ground state with (a) the bond dimension $\chi=1000$ and the periodic dimension $l_2 = 9$, and (b) the bond dimension $\chi=1200$ and the periodic dimension $l_2 = 6$. This comparison shows that the correlated pooling scheme can precisely detect the first-order phase transition (vertical red line), whereas topological states close the phase transition are misclassified for uncorrelated pooling, see the inset. }
    \label{fig:stackedSquare}
\end{figure*}

To demonstrate the advantage of our QCNN, we consider incoherent Pauli-$Y$ errors that simultaneously flip the parity of two plaquette and two star operators, introducing correlations between them. The error threshold for Pauli-Y errors is considerably increased to $3.48\%$ for correlated pooling compared to  $2.28\%$ for uncorrelated pooling, see the supplemental material \cite{SM}. This shows that the error correction capability is boosted by taking correlations into account.

In addition to incoherent Pauli-Y errors, we assume a coherent magnetic field in the $Y$ direction that drives the Toric Code across a first-order phase transition into the trivial phase \cite{Dusuel11}. We plot the output for uncorrelated and correlated pooling in Figures~\ref{fig:stackedSquare}a and \ref{fig:stackedSquare}b for an error probability $p_Y=3\%$. The output of the QCNN with correlated pooling is larger for all states in the topological phase than the output for uncorrelated pooling. Due to the limited ability to correct correlated errors, uncorrelated pooling misclassifies topological states close to the phase boundary as the output for $d=1,2$ is smaller than for $d=0$, see the inset of \autoref{fig:stackedSquare}a. In contrast, the QCNN with correlated pooling correctly recognizes all topological states. Hence, by exploiting correlations between measurement outcomes, the QCNN correctly recognizes the topological phase in the presence of strong correlated errors close to the phase transition, where classical approaches misclassify topological states.  

\begin{figure}
    \centering
    \includegraphics[width=\columnwidth]{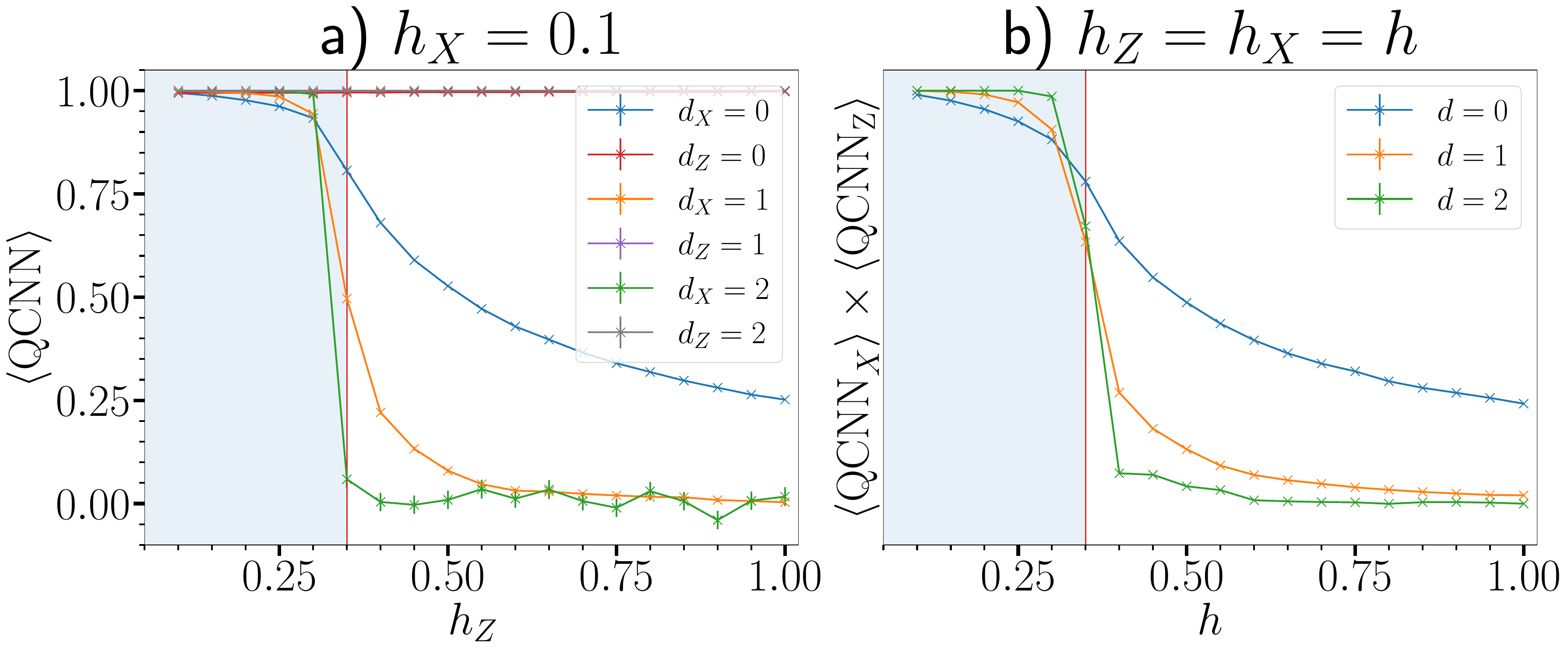}
    \caption{a) Output of the QCNN for the MPS samples under a variation of $h_Z$ in the presence of a weak magnetic field in X-direction with $h_X=0.1$ with the bond dimension $\chi=1250$. The output is displayed for simultaneous pooling in the X(Z)-basis for the corresponding QCNN depth $d_X$($d_Z$). The phase transition for $h_Z$ is correctly recognized in the presence of a field in X-direction. Furthermore, the output of the pooling layers $d_Z$ quickly converges to unity due to the weak $h_X=0.1$. b) Detection of the multicritical point \cite{vidalLowenergyEffectiveTheory2009} in the presence of X- and Z-field with $h=h_X=h_Z$ for the bond dimension $\chi=1500$. We take the product of the pooling in the X- and Z-basis for each QCNN layer to show the combined phase recognition along the multicritical line $h_X=h_Z$. A split plot can be found in the supplemental material \cite{SM}.}
    \label{fig:simultaneous_field}
\end{figure}

Finally, we test the QCNN on ground states of $H_\textrm{mag}$ with magnetic field simultaneously acting in both X- and Z-direction. First, we consider constant $h_X = 0.1$ leading to a non-vanishing probability of error syndromes $B_i = -1$ on vertices and thus $M_{\mathrm{QCNN}}^0<1$, see \autoref{fig:simultaneous_field}a. Second, we consider the transition into the trivial phase driven by perturbations in both X- and Z-directions for $h_X=h_Z = h$ \cite{vidalLowenergyEffectiveTheory2009}. The QCNN output rapidly decreases with increasing depth $d$ above the multicritical point $h=0.35$, see \autoref{fig:simultaneous_field}b. This provides clear evidence that the QCNN recognizes the phase transition also for these cases.

In conclusion, we generalized QCNNs to two-dimensional systems and showed that they recognize the topological phase of the Toric Code. The QCNN output is robust against errors below a threshold probability allowing for their realization on current quantum computers under NISQ conditions.  In processing quantum states via the QCNN quantum circuit, correlations between all stabilizer elements of the Toric Code are taken into account. As a result, the threshold probability for correlated errors is increased compared to classical approaches based on direct measurements as in \cite{congEnhancingDetectionTopological2024}. Therefore, the QCNN correctly identifies the topological phase in the presence of strong correlated errors, where classical approaches misclassify topological states. Furthermore, all pooling layers can be efficiently performed in classical post-processing to facilitate the QCNN implementation with restricted circuit depths. In future research, one can further investigate the error correction procedure in the pooling layers to increase the error threshold and enhance the phase detection by a faster convergence to a step function with the number of pooling layers. We expect that our work motivates the exploitation of QCNNs for the characterization of less-understood quantum phases of matter such as symmetry-enriched topological order \cite{mesarosClassificationSymmetryEnriched2013} and quantum spin liquids \cite{savaryQuantumSpinLiquids2016}. Implementing the pooling layers as quantum circuits is an interesting possibility for characterizing such complex systems with reduced sample complexity, and allowing for the parametrization and training of the QCNNs \cite{caroGeneralizationQuantumMachine2022, liuModelIndependentLearningQuantum2022, pesahAbsenceBarrenPlateaus2021}.

\begin{acknowledgments}
We thank Michael Knap, Frank Pollmann and  Kai Phillip Schmidt for helpful discussions and feedback.
This work is part of the Munich Quantum Valley, which is supported by the Bavarian state government with funds from the Hightech Agenda Bayern Plus. Furthermore, this work was supported by the EU program HORIZON-MSCA-2022-PF project 101108476 HyNNet NISQ (PZ) and by the Alexander von Humboldt Foundation (NAM).
\end{acknowledgments}
\nocite{hauschildEfficientNumericalSimulations2018}

\bibliography{lib_truncated}

\appendix

\onecolumngrid

\newpage

\section{Supplemental Material}
\subsection{Explicit Structure of the QCNN}
In this section, we explain the explicit structure of our QCNN in full detail. We start by discussing the overall structure of the QCNN. The QCNN is a phase recognition circuit, which is an analog of classical Convolutional Neural Networks (CNNs).
The QCNN can be written as a unitary of the form \cite{congQuantumConvolutionalNeural2019,herrmannRealizingQuantumConvolutional2022,lakeExactQuantumAlgorithms2022,zapletalErrortolerantQuantumConvolutional2023},
\begin{equation}
	U_\mathrm{QCNN}=U_\mathrm{FC} U_\mathrm{CP}^{(d)}...U_\mathrm{CP}^{(2)}U_\mathrm{CP}^{(1)} \, ,
\end{equation}
where the unitary $U_\mathrm{CP}^{(j)}$ represents a combination of a convolutional and a pooling layer, $d$ is the number of subsequent layers and the unitary $U_\mathrm{FC}$ is a fully connected layer that prepares the state of the output qubits, which are subsequently measured.
For the $j^\mathrm{th}$ layer in the QCNN structure, we specifically define
the unitary $U_\mathrm{CP}^{(j)}$ as $U_\mathrm{CP}^{(j)}=\tilde{U}_\mathrm{P}^{(j)}
	(U_\mathrm{C}^{(j)})^{\dagger}U_\mathrm{C}^{(j-1)}$,
where $\tilde{U}_\mathrm{P}^{(j)}$ is the unitary implemented by the pooling
procedure, $(U_\mathrm{C}^{(j)})^{\dagger}$ is the adjoint of the convolution of the current layer and
$U_\mathrm{C}^{(j-1)}$ is the convolution of the preceding layer
$j-1$.

If we construct the
convolutional layers from Clifford gates and the pooling procedure via single qubit Pauli-operations that are controlled by the results of measurements of other qubits, we can shorten the QCNN circuit dramatically by pushing $U_\mathrm{C}^{(j)}$ through the pooling circuit $\tilde{U}_\mathrm{P}^{(j)}$ \cite{herrmannRealizingQuantumConvolutional2022,zapletalErrortolerantQuantumConvolutional2023a}. We then obtain a modified pooling layer $U_\mathrm{P}^{(j)}$ and subsequently cancel all convolutions except the first one, such that we can write the QCNN unitary as
\begin{equation}
	U_\mathrm{QCNN}=U_\mathrm{FC} 
	U_\mathrm{P}^{(d)}...U_\mathrm{P}^{(2)}U_\mathrm{P}^{(1)} U_\mathrm{C}^{(0)}\,.
 \label{eqQCNN}
\end{equation}
As a consequence, only the initial convolutional layer $U_\mathrm{C}^{(0)}$ is executed on the quantum device, while all other layers can be computed in classical post-processing. We define $U_C = U_\mathrm{C}^{(0)}$ for brevity.

For the convolutional layer, the starting point is the generation of the reference state. The intuitive choice for the reference state is the ground state of the unperturbed Toric Code, $H_{\text{TC}}$. From \cite{satzingerRealizingTopologicallyOrdered2021} we know that there is an exact circuit that can prepare the intrinsic topological order of the Toric Code from an all $\ket{0}$ initialization of the qubits. This is achieved via the preparation of the plaquette stabilizers. We start by choosing one representative qubit for each plaquette. Then, we bring the representative qubits to the $\ket{+}$ state with a Hadamard gate and subsequently entangle the other qubits with CNOT gates that are controlled on the representative qubit. This is sufficient to implement the X-parity between the four qubits of each plaquette. 

\begin{figure}
\includegraphics[width=0.7\columnwidth]{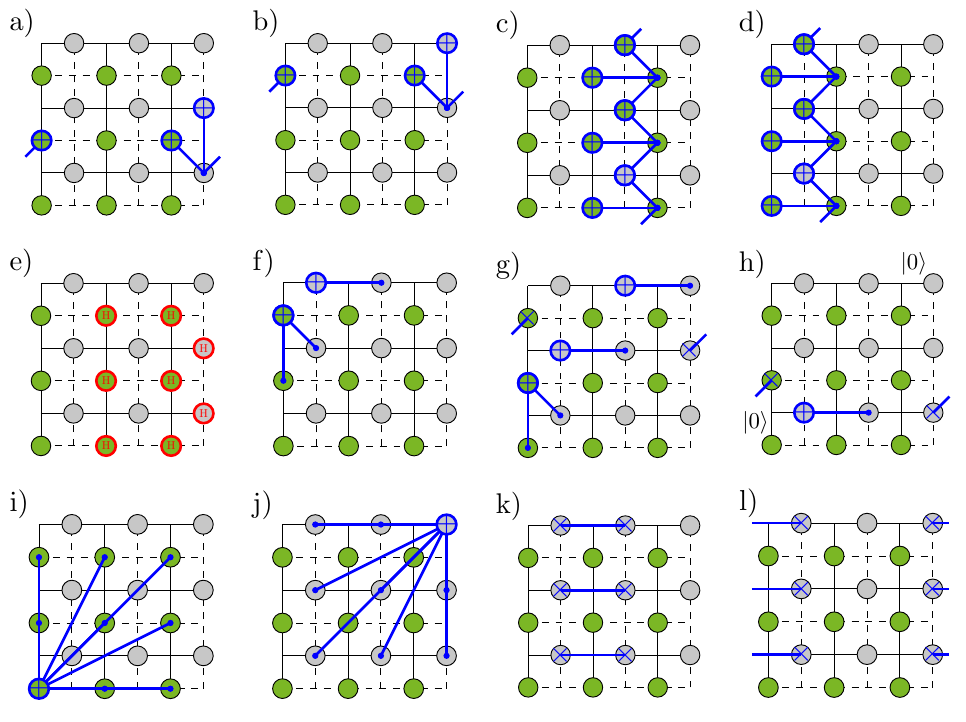}
	\caption{Sequence of all gates in the convolution $U_\mathrm{C}^{(0)}$ (see main text) for a $3\times 3$ lattice with periodic boundaries. Figures a) to e) depict the inverse Toric Code ground state preparation algorithm $U^\dagger _\mathrm{prep}$. The blue gates are CNOTs, for which the $+$ mark the respective targets and the circles mark the controlling qubits. Here we show how for the rightmost column of the lattice the representative qubits are the lower gray qubits, whereas for all other columns of plaquettes, the right green qubit is the representative qubit. This is an example of a set of representative qubits that ensures during the ground state preparation circuit $U_\mathrm{C}$ that the representative qubits have not yet been entangled before their plaquette is prepared. On lattice e) we also visualize the Hadamard gates (red), which also are part of the inverse preparation circuit. Figures f) to h) show additional CNOTs that are necessary to map the vertex stabilizers correctly. These gates do not act on the mapping of the plaquette stabilizers as they are only controlled on gray qubits on the horizontal lattice edges. In g) and h) we additionally implement SWAP gates, denoted by the cross symbol $\times$, over the periodic lattice edges to correct the broken translational symmetry in a) and b) from choosing gray qubits as representative qubits. Lastly, in figures h) to j) we reset the bottom left and top right qubits, before mapping the product of all other stabilizers of the corresponding basis to the qubits. Before the reset, the procedure of a) to g) mapped a logical X and a logical Z operator to these two qubits, respectively. In figures i) and j) specifically, we apply CNOT gates that target the qubit in the corner and are controlled by all other qubits of the corresponding sublattice. With this step we remove all information about the logical subspace, but ensure that all stabilizer generators are mapped to the correct sites and thus can apply the local pooling operations in a translationally invariant way. Lastly, in figures k) and l), we swap each column of qubits on horizontal lattice edges one step to the right, such that syndromes of Pauli-Y errors on the lattice create errors on four qubits in a diamond shape. This is only necessary as a preparation for the pooling scheme if there are correlated errors that affect both horizontal and vertical lattice edges simultaneously.}
	\label{figs_convolution}
\end{figure}

The representative qubits need to be chosen in a sequence that ensures that they have not been entangled with other qubits before. Throughout the bulk of the lattice, this method is easy to apply via the preparation of plaquettes in sequential columns. Only at the periodic boundaries it is important to be mindful of choosing the correct representative qubits, that have not yet been entangled by the first column of plaquettes. As neighboring vertex and plaquette stabilizers always share two qubits, this circuit also satisfies the requirement of the Z-parity between the qubits of each vertex stabilizer and therefore, the Toric Code ground state is correctly prepared. We will call this circuit $U_\mathrm{prep}$, see \autoref{figs_convolution}.

Accordingly, with $  U^\dagger_\mathrm{prep}\ket{\psi_\mathrm{toric}}=\ket{0}^{\otimes N}$
we know that the inverse preparation circuit $U^\dagger_\mathrm{prep}$ takes the Toric Code ground state back to the all-zero state. If we apply the inverse preparation circuit $U^\dagger_\mathrm{prep}$ to a perturbed lattice state $\ket{\phi}\neq\ket{\psi_{toric}}$, the circuit will map the parity of all plaquettes to the representative qubits. 

In \autoref{figs_convolution} we show the full sequence of gates for the convolution $U_\mathrm{C}^{(0)}$ as in the main text for a small $3\times 3$ example. This includes additional CNOT and SWAP gates that we need to add to $U^\dagger_\mathrm{prep}$ in order to also map vertex stabilizers to representative qubits, as we have already done with the plaquettes. During this procedure, we are also removing the mapping of the logical operators by resetting two qubits, see \autoref{figs_convolution}h. Hence, we are losing all information about which state within the subspace of the degenerate ground states of the Toric Code was put in and the QCNN accordingly only preserves information about the topological phase.

\begin{figure}[!h]
	\includegraphics[width=0.7\columnwidth]{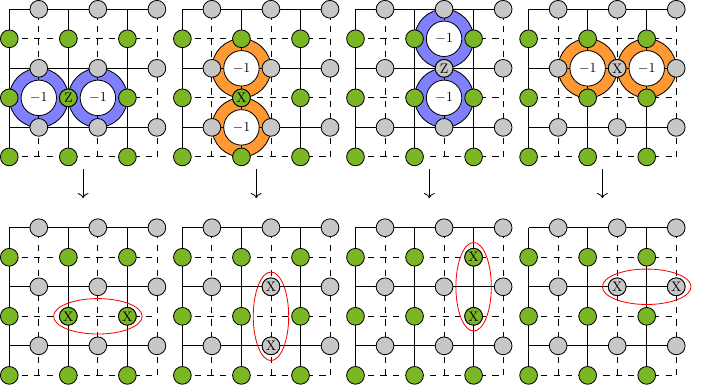}
	\caption{Effect of single-qubit Pauli-Z and Pauli-X errors. In the top row, we show the corresponding error syndromes. A Pauli-Z error will flip the parity of adjacent plaquettes (X-stabilizers), while a Pauli-X error will induce a syndrome on neighboring vertices (Z-stabilizers). The lattices in the bottom row depict the mapping of the error syndromes to error pairs on individual qubits under the convolution. All plaquettes are mapped to qubits on vertical lattice edges (green) and the vertices are mapped to qubits on horizontal lattice edges (gray).}
	\label{syndrome_evo}
\end{figure}

After the implementation of the convolution, we now outline the structure of the pooling layer. The advantage of mapping the stabilizer measurements to individual qubits is that perturbations on the Toric Code can be easily revealed. Single-qubit perturbations induce error syndromes on the lattice of the Toric Code in its error-correcting functionality. As every qubit is part of four stabilizer elements, a single-qubit Pauli error will induce a negative parity on the neighboring stabilizer elements that anticommute with the error. In our case, such a single-qubit error will be converted to a pair of Pauli-X errors on neighboring qubits after the convolution. We show this property in \autoref{syndrome_evo}.

Hence, the challenge is to implement an error correction scheme that exploits the correlated error pairs generated by the convolution $U_\mathrm{C}^{(0)}$. We choose a set of target qubits $\{t_j\}$ that are error-corrected and propagated to the next layer. In general, we want to remove short-range behavior and evaluate only global topological order on the finally remaining output qubits. As in quantum error correction, one needs to find a procedure that is capable of removing any singular error and reducing the size of error clusters relative to the total number of qubits. By singular error, we here mean that it is an error that is sufficiently far away from all other errors such that they do not interact. On the other hand, an error cluster can mean directly neighboring or overlapping error patterns after the convolutional layer. 

In this work, we propose to choose the set of target qubits $\{t_j\}$ to be evenly spaced out on the lattice with a periodicity of three. This ensures that each target qubit $t_j$ has four neighboring qubits that are at the same time not nearest neighbors of another target qubit. Therefore, singular errors that are detected on these neighboring qubits can be unambiguously associated with the specific target qubit. After the convolution, this allows for the correction of all error pairs that originate from singular errors on the input and are located on the target qubit via the application of CNOT gates controlled on the four nearest neighbors of the target qubit. We call these the control qubits $\{c_k\}= \mathcal{N}(t_j)$ with $\mathcal{N}(t_j)$ referring to the neighborhood of $t_j$. However, error pairs located between the target qubits will propagate onto the target qubits and thus remain on the renormalized lattice. Therefore, it is necessary to apply Toffoli gates in an additional step to the target qubits that are controlled by the control qubits $\{c_k\}$ and by their respective nearest neighbors without the target qubit $\{n_l\}=\mathcal{N}(c_k)\backslash t_j$. This will then remove a wrongful correction that was introduced by the CNOT gates. According to the chosen set of target qubits $\{t_j\}$ we can calculate the required number of qubits in each layer $l$ for a specific depth $d$ of the QCNN as
\begin{equation}
    N_\mathrm{qubits}(l)=2\times 3^{2(d-l)}\,.
\end{equation}
The pooling is shown for a small example in \autoref{fig:poolingToffolis} and can be expressed as a unitary as
\begin{equation}
U_{P}= \prod_{c\in
		\mathcal{N}(t)}\left(\prod_{n\in \mathcal{N}(c)\backslash t} \textrm{Toff}_{nct}\right)
	\textrm{CNOT}_{ct}\,,     
\end{equation}
whereas $\mathcal{N}$ describes the set of nearest-neighbors. We can also express a Pauli-Z on a target qubit, which is the observable of interest after the pooling, via the following transformation under a single pooling layer
\begin{align*}
	U_{P}^\dagger Z_t U_{P}= & \left(\prod_{c\in
		\mathcal{N}(t)}\prod_{k\in \mathcal{N}(c)\backslash t} \textrm{Toff}_{kct}
	\textrm{CNOT}_{ct}\right)Z_t \left(\prod_{d\in \mathcal{N}(t)}\prod_{l\in
	\mathcal{N}(c)\backslash t}\textrm{CNOT}_{dt}\textrm{Toff}_{ldt}\right)=                      \\
	=                            & \left(\prod_{k,c}\textrm{Toff}_{kct}\right) \left(\prod_{c}
	\textrm{CNOT}_{ct} \right)Z_t \left(\prod_{d} \textrm{CNOT}_{dt}\right)
	\left(\prod_{l,d}\textrm{Toff}_{ldt}\right)=                                                  \\
	=                            & \left(\prod_{k,c}\textrm{Toff}_{kct}\right)\left(\prod_{c} Z_c
	\right)Z_t\left(\prod_{l,d}\textrm{Toff}_{ldt}\right)=                                        \\
	=                            & \left(\prod_{c} Z_c \right)
	\left(\prod_{k,c}\textrm{Toff}_{kct}\right)Z_t\left(\prod_{l,d}\textrm{Toff}_{ldt}\right)=    \\
	=                            & \left(\prod_{c\in  \mathcal{N}(t)} Z_c\right) Z_t
	\left(\prod_{\substack{c\in \mathcal N(t),                                                    \\k\in \mathcal{N}(c)\backslash t}}
	\textrm{CZ}_{kc}\right),
\end{align*}
where we have used the identities
\begin{align*}
	\textrm{CNOT}_{ct}Z_t\textrm{CNOT}_{ct}   & =Z_c Z_t ~~\mathrm{and}                                            \\
	\textrm{Toff}_{kct}Z_t\textrm{Toff}_{kct} & = (
	\ket{11}\bra{11}_{kc}\otimes X_t + (1-\ket{11}\bra{11}_{kc})\otimes
	\mathds{1}_t) Z_t \textrm{Toff}_{kct}                                                                          \\
	                                          & =-\ket{11}\bra{11}_{kc}\otimes Z_t+(1-\ket{11}\bra{11})\otimes Z_t \\
	                                          & =Z_t(1-2\ket{11}\bra{11})                                          \\
	                                          & = Z_t \textrm{CZ}_{kc}\,.                                             
\end{align*}
In general, this pooling procedure as part of the QCNN can be written and implemented as a quantum circuit. However, as it only consists of controlled Pauli gates that act on Pauli strings followed by measurements, one can also perform these operations as classical post-processing on measurement snapshots after having performed the convolutional layer. 

In this work, we propose two different pooling schemes. In the first case, as shown in \autoref{fig:poolingToffolis}a and b, X-stabilizers and Z-stabilizers are separately processed in the pooling procedure. As a result, the convolutional layer can be replaced by the direct sampling of the ground states in both the Z-basis and X-basis. For each measurement snapshot in the Z-basis (X-basis), we can determine all Z-stabilizer (X-stabilizer) values. In comparison to the measurement after the convolutional layer, the direct measurement of the input state in both Z-basis and X-basis increases sample complexity by a factor of two.

\begin{figure}
    \centering
    \includegraphics[width=8.6cm]{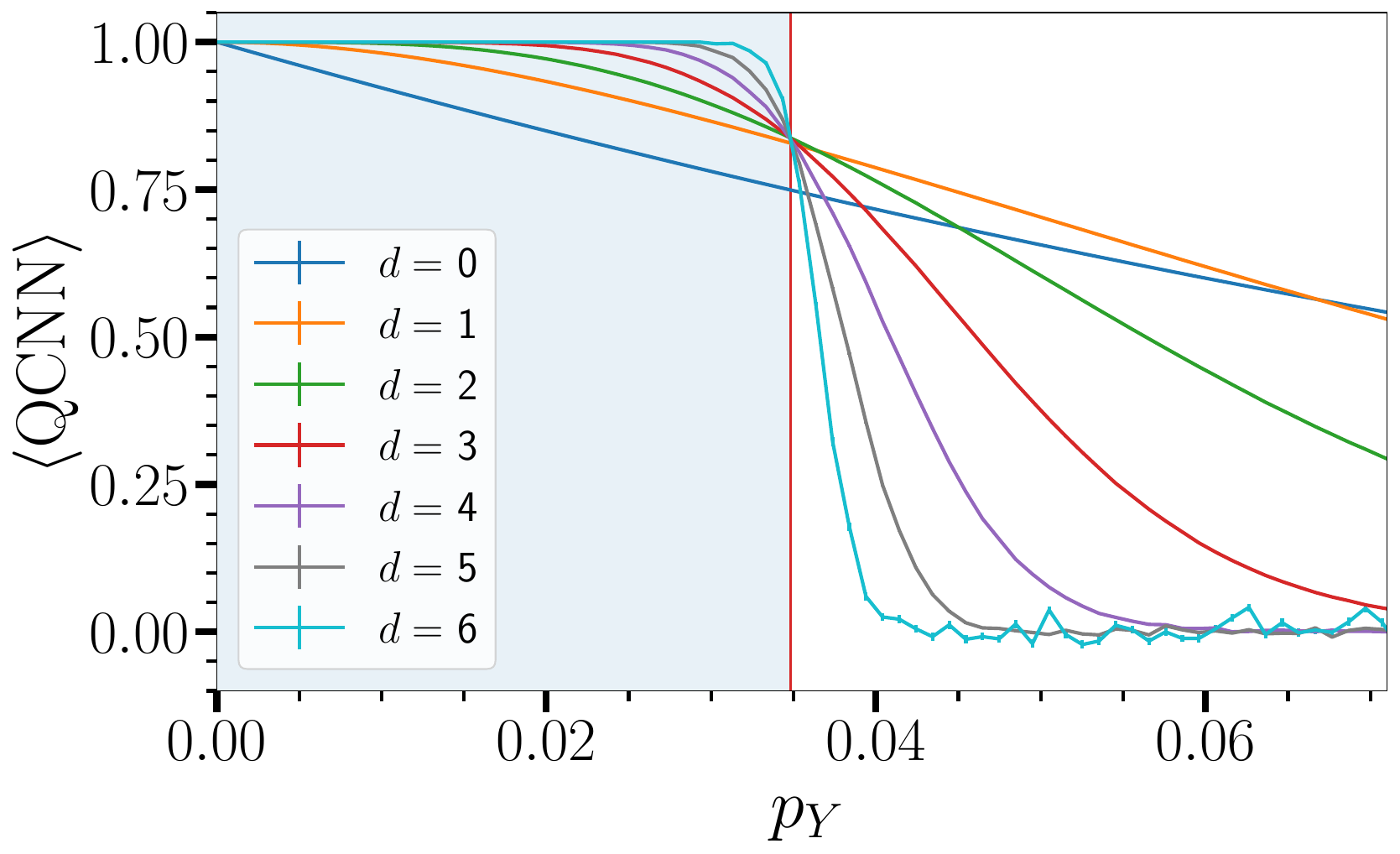}
    \caption{QCNN output for incoherent Pauli-Y errors with correlated pooling. Specifically, we choose the pooling procedure depicted in \autoref{fig:poolingToffolis}c and d for the first layer ($d=1$) and then return to the former pooling scheme as in \autoref{fig:poolingToffolis}a and b for all following layers ($d\geq 2$). This is necessary as the correlated pooling scheme either removes the four-qubit error syndromes or maps overlapping errors to pairs on the renormalized lattice. However, the greater ratio of retained and corrected target qubits from the first pooling layer then boosts the total error threshold to $p_{\mathrm{th},Y} =3.48\%$. }
    \label{fig:PauliStackedPooling}
\end{figure}

\begin{figure*}
	\includegraphics[width=\columnwidth]{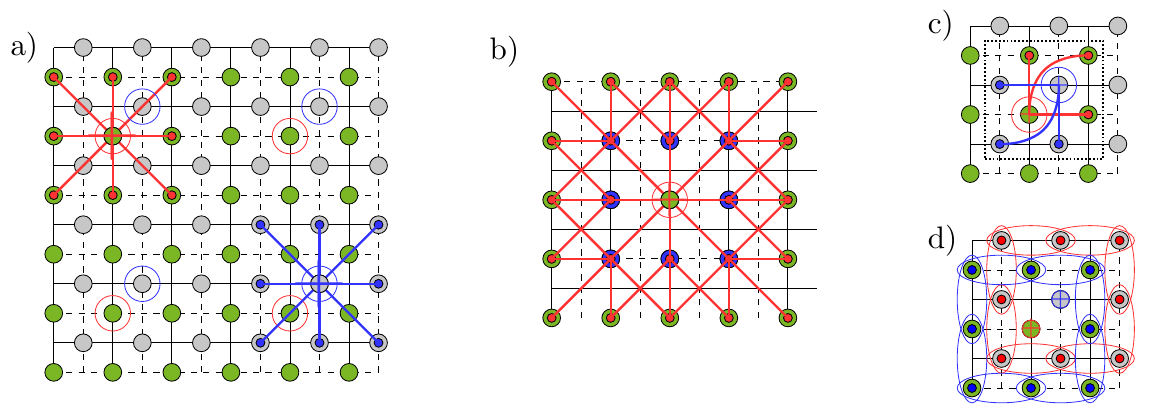}
	\caption{Graphical depiction of gates in the pooling procedure. Green and gray qubits form the two shifted sublattices of the Toric Code. Encircled are the target qubits, that will be propagated to the higher layer. The red circles mark the target qubits for the sublattice of the qubits on vertical edges (pooling in the X-basis), whereas the blue circle marks the target qubits on horizontal lattice edges (pooling in the Z-basis). a) On the target qubits, we apply CNOT gates controlled by their nearest neighbors on their respective sublattices. For visibility reasons, we only show the CNOT gates for one target qubit of each sublattice. b) Here, we show only the sublattice of qubits on vertical lattice edges. Subsequently to the CNOT gates in a), we apply Toffoli gates controlled on the nearest and next-nearest neighbors of the target qubits. This ensures that pairs of errors that are anchored on the nearest neighbors but not on the target qubit do not cause a wrong flip of the target qubit. Specifically, such an error pair first triggers a flip of the target qubit via the CNOTs, which is then consequently removed after the application of the Toffolis. c) With an alternative set of CNOT gates, we can achieve a higher ratio of target qubits for the correlated pooling scheme. Compared to propagating one of every ninth qubit with the uncorrelated pooling scheme, we can correct one out of four qubits in one pooling layer. This is applicable under Pauli-Y perturbations via the exploitation of the larger support of the corresponding syndromes compared to syndromes of X-/Z-errors. As the convolutional layer maps the syndrome of a single Y-error to a diamond pattern on four qubits, we can use the available additional information for more effective error correction. The dotted rectangle marks the edges of the unit cell, which generates the set of target qubits, when iterated over the lattice. d) Contrary to the uncorrelated pooling scheme, the control and target qubits of the Toffoli gates lie at different sublattices, allowing for the correction of correlated errors. In this scheme, ovals indicate pairs of qubits that are the controls of a Toffoli gate acting on the target qubit with the corresponding color, thus coupling green and gray qubits. In combination with the CNOT gates, this scheme can correct a single Y error at an arbitrary qubit.}
	\label{fig:poolingToffolis}
\end{figure*}

We now discuss a second pooling scheme, which is beneficial if the perturbation of the Toric Code is formed by Pauli-Y errors. This procedure aims to exploit the larger support of the syndrome of a Y-error, as it flips the parity of the adjacent X- and Z-stabilizers, forming an error syndrome in a diamond shape. As a result, the error syndromes on plaquettes and vertices are correlated.  We can use these correlations to boost the error correction rate, i.e., the ratio of correctable error syndromes and the number of qubits discarded in each pooling layer. To this end, we design a new correlated pooling layer depicted in \autoref{fig:poolingToffolis}c and \autoref{fig:poolingToffolis}d. In contrast to the uncorrelated pooling layer that required discarding eight out of nine qubits, the correlated pooling layer discards only three out of four qubits, retaining a larger number of target qubits. Thanks to the correlations between the parity flips of X stabilizers and Z stabilizers, we can correct the syndrome of a single Y error at an arbitrary qubit by performing the CNOT and Toffoli gates shown in \autoref{fig:poolingToffolis}c and \autoref{fig:poolingToffolis}d. As a larger number of qubits is retained in each layer, the error correction rate is increased.
 As we show in \autoref{fig:PauliStackedPooling}, this boosts the error threshold for Pauli-Y errors considerably to $3.48\%$ from the former $2.28\%$. 

To make use of this pooling scheme, it is necessary to apply the convolution as a quantum circuit and thus acquire a correlated snapshot of both stabilizer bases after measuring all qubits. One can then apply the pooling again in classical post-processing. Specifically, we also need to apply the additional SWAP-gates \autoref{figs_convolution}k and l during the convolution, which are not necessary for the uncorrelated pooling. 

\subsection{Details on the Matrix-Product-State Simulations}
\label{mps_details}
\begin{figure}
    \centering
    \includegraphics[width=8.6cm]{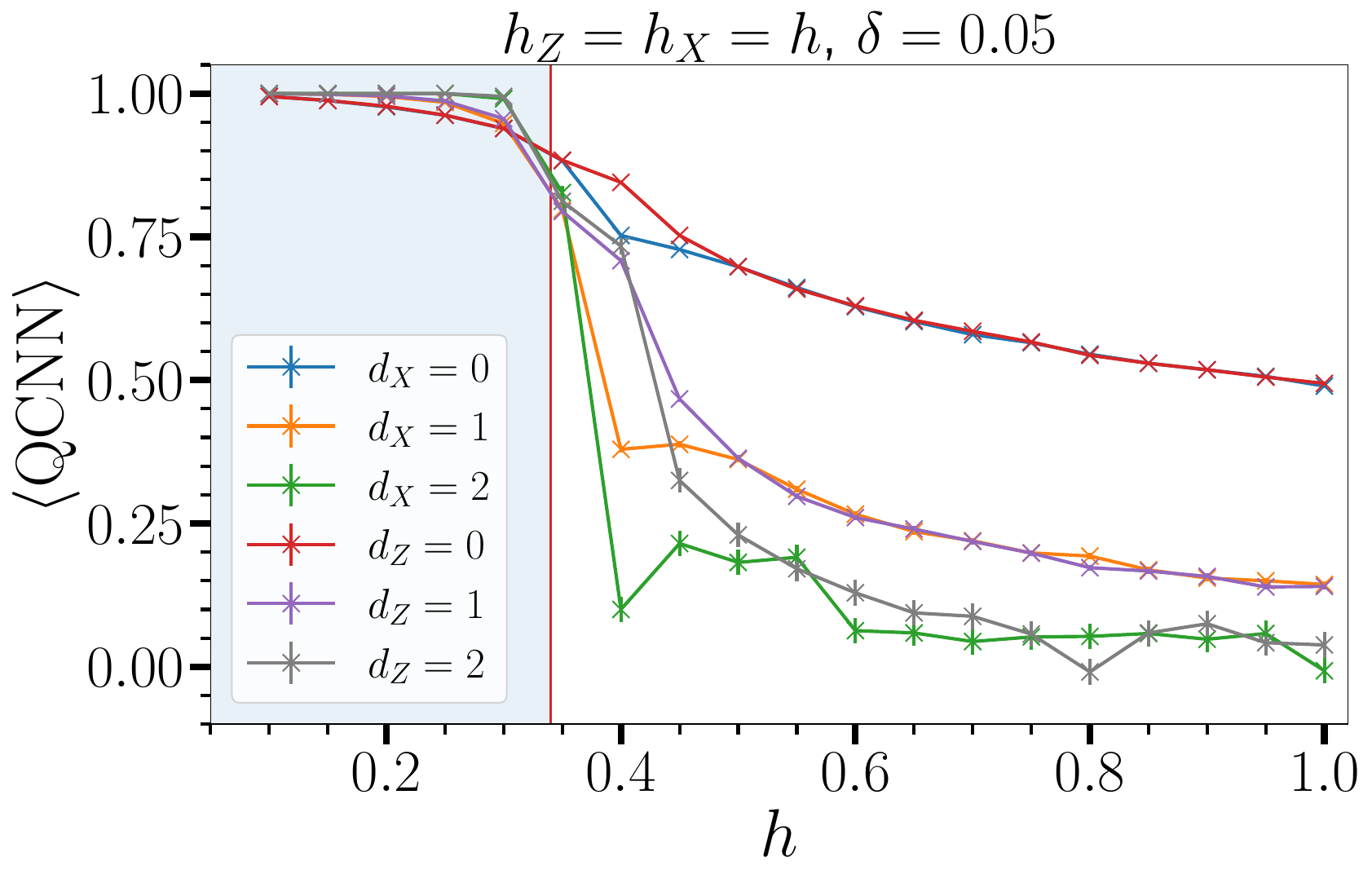}
    \caption{QCNN output for the multicritical point along $h_X=h_Z$. In comparison to the figure in the main text, we plot each pooling layer individually instead of taking the product over the pooling in both bases. We observe a splitting in the values of corresponding $d_X$ and $d_Z$. This results from the property that the ground state of the Toric Code manifold in this case is two-fold degenerate with a respective bias towards one type of stabilizer. If we initialize the MPS with $\delta=-0.05$, the behavior of both pooling bases will interchange.}
    \label{fig:multi_discrete}
\end{figure}

In this section, we provide additional information on our implementation of the MPS. We use the TenPy python library \cite{hauschildEfficientNumericalSimulations2018} to initialize the MPS on an infinite cylinder geometry. Accordingly, the horizontal dimension $l_1$ is infinite with open boundaries and the vertical dimension $l_2$ is periodic. As a strategy to generate the MPS we run the infinite Density Matrix Renormalization Group (iDMRG) algorithm \cite{vidalClassicalSimulationInfiniteSize2007} in two distinct steps in order to consistently select a specific ground state of the degenerate manifold. In the first step, we add an energy penalty for the logical operators to select the +1 eigenstate of the Wilson and t'Hooft loop operators that span the periodic dimension of the cylindric lattice before we optimize without the penalty in a second iteration of iDMRG. This ensures that we avoid convergence to superpositions of the different ground states. We can then numerically compute ground states with a bond dimension of up to $\chi=2000$.  However, we restrict the bond dimension to $\chi=1250$ for most plots in this manuscript, as we found that increasing it further did not lead to a visible change in our findings. Special attention is also required for the MPS initialization in the case of $h_X=h_Z$ as in the main text. As the ground state under this condition is two-fold degenerate \cite{vidalLowenergyEffectiveTheory2009}, we first run iDMRG with $h_X=h-\delta $ and $h_Z=h+\delta$ for a small $\delta=0.05$ in order to select the corresponding eigenstate. In a second run of iDMRG, we then optimize the MPS for $h_X=h_Z$ to converge to the correct ground state with the previous run as initialization. Note that, for the other ground state obtained with $\delta = -0.05$, the QCNN output is similar to \autoref{fig:multi_discrete} but the values in X- and Z-basis interchanged.

As we discussed in the previous section, the QCNN can be entirely run as a quantum circuit but has the benefit that {for uncorrelated pooling, it can also be fully realized in classical post-processing. We sample spin configurations in the Z-basis and in the X-basis from the ground states. We determine the QCNN output in classical post-processing as explained in the previous section. For our simulations, we draw samples for a total of 486
qubits on a $2 l_1 l_2 = 2 \times 27 \times 9$ qubit lattice. A reduced schematic of the lattice can be found in \autoref{MPS_schematic}. In the lattice dimensions, the factor of two refers to the two shifted sublattices for qubits either on horizontal or vertical lattice edges. Furthermore, $l_1$ is the number of qubits in the horizontal dimension and $l_2$ in the vertical
direction.

We chose these specific dimensions $l_1$ and $l_2$ such that we can apply two consecutive pooling layers, which fixes the vertical (periodic) dimension to $l_2=9$ qubits. Intuitively, the same minimum number of qubits should also suffice for the QCNN in horizontal dimension, but as the MPS is non-periodic in this direction, we have to use additional qubits as a buffer. This is necessary because the open boundaries break the properties of the convolution $U_C^{(0)}$ at the edges. We solve this problem by first applying the QCNN to all sampled qubits, but only extracting information from the qubits, for which the light cone is restricted to the bulk of the lattice and does not span out to the edges as sketched in \autoref{QCNN_dim4}. This requirement must be respected in every layer of the QCNN. Hence, we can find a lower limit for the required number of qubits by defining the lattice configuration for the highest layer. From the QCNN
structure, we know that the number of qubits increases by a factor of 9 for every pooling layer and the highest output layer consists of a column of two qubits. To mitigate the edge interactions on the non-periodic dimension of the cylinder code, we need to trim at least one column of qubits on each side of the contributing qubits in horizontal dimension. Accordingly, the output layer must be grown to a number of $N=3\times 2$ qubits to the triple of the original size in dimension $l_1$. As all qubits in the higher layers must be propagated from lower layers, the lower layers must all also be tripled in horizontal dimension to a final size of $l_1=27$, but all target qubits but the left- and rightmost column can contribute to the results.
\begin{figure}
\includegraphics[width=\columnwidth]{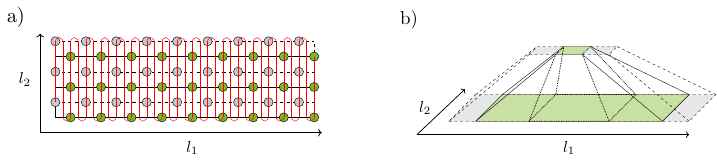}
	\caption{a) Structure of the MPS on the cylindric lattice. Using the snake over the lattice we map the system to a one-dimensional MPS. In this geometry, qubit interactions span the periodic dimension $l_2$, whereas the lattice goes to infinity in the dimension $l_1$, forming an infinite cylinder. b) Lattice dimensions for MPS simulations. The green area from top to bottom resembles the light cone of the output qubits. Due to the infinite horizontal dimension, we need to ensure to only evaluate bulk behavior in horizontal direction in order to avoid propagating effects of the wrong boundary conditions to the output qubits. Therefore, we need to triple the horizontal lattice dimension in every layer, such that the edge behavior can be mapped in the highest layer to qubits that do not need to be measured. In the figure, the tripled dimension is shown with the outer dashed lines. In comparison, the inner dashed lines show a minimal example of the light cone of the output qubits. This resembles the minimum reduction factor of qubits that are necessary for one pooling step. To get improved statistics in each layer versus the minimal example we can evaluate all qubits up to the edge in the green region.}
        \label{MPS_schematic}
	\label{QCNN_dim4}
\end{figure}

For correlated pooling, the convolutional layer $U_C$ must be implement as a quantum circuit. To this end, we exploit that the Hamiltonian $H_{\rm mag}=\sum_{j=1}^{N_P}c_j P_j$ is a sum of Pauli strings $P_j$ with real coefficients $c_j$ and the convolutional layer is a Clifford circuit. As a result, we can analytically express the transformed Hamiltonian $H_{C} = U_C H_{\rm mag} U_C^{\dagger}$ as a sum of transformed Pauli strings $P_j^C=U_C P_j U_C^{\dagger}$ using the following gate identities $X_j=H_j Z_j H_j$, $\textrm{CNOT}_{i,j}X_i\textrm{CNOT}_{i,j} =X_i X_j $ and $\textrm{CNOT}_{i,j}Z_j\textrm{CNOT}_{i,j} =Z_i Z_j $. The ground state $|\psi_C\rangle= U_C |\psi_{\rm mag}\rangle$ of $H_C$ is the ground state $|\psi_{\rm mag}\rangle$ of $H_{\rm mag}$ transformed by the convolutional layer $U_C$. We use the iDMRG algorithm to find the transformed ground states $|\psi_C\rangle$ for the infinite cylinder with the periodic dimension $l_2=6$, the bond dimension $\chi=600$, varying field strengths $h_Y$, and $h_x=h_Z=0$. To benchmark numerical errors, we use that the unitary transformation $U_C$ does not change the ground state energy. The numerical error in energy per qubit compared to the iDMRG simulation of the original Hamiltonian $H_{\rm mag}$ with the bond dimension $\chi=300$ is $|\Delta E| < 1.2\cdot 10^{-3}$ for all field strengths $h_Y$. Furthermore, we observe that increasing the bond dimension to $\chi=1200$ does not lead to a visible change in our results.

A special attention has to be paid to the critical value $h_Y=1$ of the field strength, where the ground state is two-fold degenrate. We use two different initializations of the iDMRG algorithm to find both ground states. First, we initialize the iDMRG algorithm with the state with energy penalties for the Wilson and t'Hooft loops explained above. Second, we initialize the iDMRG algorithm with the paramagnetic state, which is the ground state for $h_Y\rightarrow\infty$.

Subsequently, we sample spin configurations in the Z-basis, corresponding to the plaquette and star operators of the original ground states $|\psi_{\rm mag}\rangle$. We determine the QCNN output in classical post-processing as explained in the previous section. For our simulations, we draw samples for a total of 216
qubits on a $2 l_1 l_2 = 2 \times 18 \times 6$ qubit lattice.

\subsection{Additional Numerical Simulations}

In \autoref{Pauli2D} we plot the QCNN output, for the input state being the reference state of the topological phase, over simultaneous Pauli-Z and Pauli-X noise with corresponding error probabilities $p_Z$ and $p_X$. Specifically, we apply random Pauli noise to each individual qubit in the form of the error channel
\begin{equation}
\rho=\mathcal{E}(\ket{\psi}\bra{\psi})=\sum_{l=0}^m K_l \ket{\psi}\bra{\psi}K_l^\dagger\,,
    \label{PauliNoise}
\end{equation}
where $K_l\in\{\sqrt{p_\mathds{1}}\mathds{1},\,\sqrt{p_X(1-p_Z)}X,\,\sqrt{p_X p_Z}Y, \sqrt{p_{Z}(1-p_X)}Z\}^{\otimes N}$ and $p_\mathds{1}=1- p_{X} +p_X p_{Z}-p_{Z}$. $p_{X}$($p_{Z}$) is the probability for a Pauli-X(Z) error on any single qubit. In contrast to Equation (6) of the main text, X and Z errors are uncorrelated and a Pauli-Y error occurs as the product of a Pauli-X and a Pauli-Z error on the same qubit. For this case, we consider up to 4 layers of QCNN pooling (corresponding to around 120,000 qubits). \autoref{Pauli2D} shows that the same output transition can be found
for both error bases. Again, the sharpness of the transition increases with additional QCNN layers. Furthermore, the area
classified as topological is of square shape, which is due to the symmetry under the basis transformation of the stabilizer mapping in the Toric Code. Accordingly, the corresponding stabilizer syndromes are mapped to the disjunct sublattices and processed separately in the QCNN.
\begin{figure}
	\includegraphics[width=\columnwidth]{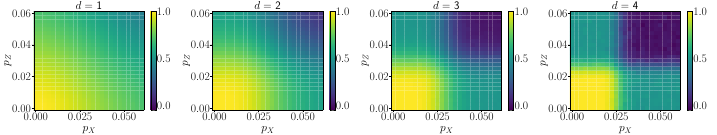}
	\caption{Two-dimensional plots for the QCNN output for simultaneous errors in the X- and Z-basis. We observe increased sharpness for the transition between the topological phase and the disordered regime for error rates higher than $p_\mathrm{th}=2.28\%$ with additional pooling layers. These plots show, that the flip errors of the respective basis do not interact with each other during the processing in the QCNN, as is apparent by the square area that is characterized as topological.}
	\label{Pauli2D}
\end{figure}

Furthermore, we tested the phase recognition capability of the
QCNN for Toric Code ground states of varying magnetic field strength, that are
perturbed by Pauli noise. To this end, we reevaluate the samples from
Figure 3 of the main text and additionally apply the Pauli noise channel (\autoref{PauliNoise}) to them.   The plots of \autoref{DMRG_robust} show that incoherent Pauli errors have, as expected, a significant influence on the QCNN output for the Toric Code ground states as soon as the error rates reach the threshold $p_\mathrm{th}$. Up to the threshold, the QCNN output behaves similarly as in Figure 3 of the main text. However, for $p_X=p_Z=2.25\%$, the output curve for layer 1 drops below the curve for layer 0. Since $\braket{M^l_\mathrm{QCNN}}$ depends on the error rate on the lattice at layer $l$, this indicates that the density of errors has increased with the first pooling layer. Such behavior is expected as the error rate is close to the threshold. In this case, the number of errors removed by the first pooling layer is smaller than the reduction in the total number of qubits and the errors can only be cleared sufficiently with the subsequent pooling layer. This behavior also starts to become apparent in the second pooling layer for larger error rates and for our last test with $p_X=p_Z=10\%$ the error density after the second pooling layer even increases beyond the density after the first layer.

The phase recognition continues to agree with the ground states in the absence of Pauli-noise up to $p_X=p_Z=3\%$, where the input states up to the phase transition are more likely to be characterized correctly since the average QCNN output $\braket{\mathrm{QCNN}}$ rests above $50\%$ in the topological phase. In our tests, we find topological input states that are falsely characterized as trivial for $p_j>3\% (j=X,Z)$ and for strong Pauli noise at $p_X=p_Z=10\%$ no input state ensemble is characterized as topological. This behavior is expected as the correlation length of the syndromes of the Pauli errors becomes so long compared to the error correction capabilities of the QCNN that the topological character of the input states remains masked.

\begin{figure}
    \includegraphics[width=0.7\columnwidth]{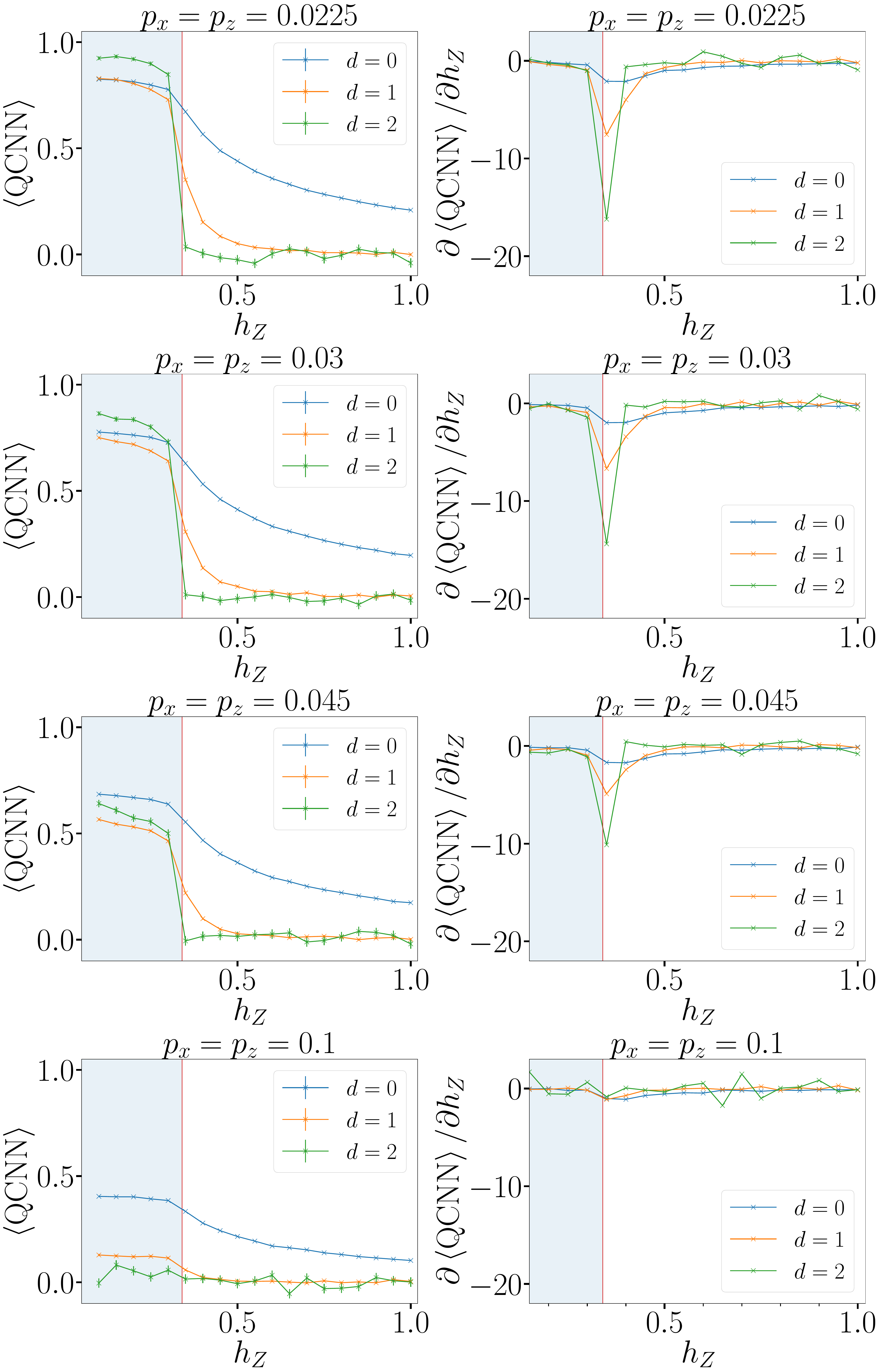}   
    \caption{QCNN output for the MPS samples over the magnetic field strength $h_Z$ under the influence of incoherent Pauli errors with rate $p_X=p_Z\in \{0.0225,\,0.03,\, 0.045,\,0.1\}$ with bond dimension $\chi=1250$. On the right side, we plot the gradient of the QCNN output to show how the increasing noise influences the detected point of steepest change. We observe that even for noise that surpasses the Pauli noise threshold $p_\mathrm{th}= 2.28\%$, the point of steepest change is still correctly identified at the phase transition $h_Z^*=0.34$ up to the resolution of our samples. Far above the error threshold, the average QCNN output no longer identifies all input states correctly. For $p_X=p_Z=0.045$ already, the output for one and two pooling layers drops below the direct measurement ($d=0$). However, the derivative of the output $\partial \braket{\mathrm{QCNN }}/\partial h_Z$ still sharply peaks at the phase transition for $p_X=p_Z=0.045$. In the bottom row, we show that for strong Pauli noise of $p_X=p_Z=0.1$, the ability to recognize the phase transition is lost.}
    \label{DMRG_robust}
\end{figure}

\end{document}